\documentclass[a4paper,twocolumn,showpacs,amsmath,pra,amssymb]{revtex4-1} 
\usepackage[T1]{fontenc}
\usepackage[latin9]{inputenc}
\usepackage{times}
\usepackage{color}
\usepackage{xspace}
\usepackage{color,ulem}
\usepackage{amssymb,amsmath}
\usepackage{amsbsy}
\usepackage{graphicx}
\usepackage{bm}
\usepackage{epstopdf}
\usepackage{float}

\newcommand{\diff}{\mathrm{d}}

\newcommand{\bx}{\ensuremath{\mathbf{x}}\xspace}
\newcommand{\x}{\mathbf{x}}

\newcommand{\bk}{\mathbf{k}}

\begin{document}
\newcommand{\nt}{\tilde{n}}
\newcommand{\Lnh}{\ensuremath{\mathcal{L}}\xspace}
\newcommand{\dbx}{\diff\bx}
\newcommand{\dbk}{\diff\bk}
\newcommand{\nbe}{\ensuremath{\bar{n}_{\mathrm{BE}}}\xspace}

\newcommand{\bro}{\mathbf{r}_1}
\newcommand{\brt}{\mathbf{r}_2}
\newcommand{\COM}[1]{\textcolor{red}{#1}}
\title{Fingerprinting Rotons in a Dipolar Condensate: Super-Poissonian Peak in the Atom-Number Fluctuations}
\author{R.~N.~Bisset} 
\author{P.~B.~Blakie}  

\affiliation{Jack Dodd Centre for Quantum Technology, Department of Physics, University of Otago, Dunedin, New Zealand.}

\begin{abstract}
We demonstrate that  measurements of atom-number fluctuations in a trapped dipolar condensate can  reveal the presence of the elusive roton excitation. The key signature is  a super-Poissonian peak in the fluctuations as  the size of the measurement  cell is varied, with the maximum occurring when the size is comparable to the roton wavelength. The magnitude of this roton feature is enhanced with temperature.
The variation in fluctuations across the condensate demonstrates that the roton excitations are effectively confined to propagate in the densest central region, realizing a density trapped   \textit{roton gas}.  While our main results are based on full numerical solutions of the meanfield equations, we also develop and validate a simple local density theory.
Finally, we consider fluctuations measured within a washer-shaped cell which filters out the contribution of modes with nonzero angular momentum  and provides a signal sensitive to individual roton modes.
\end{abstract}

\maketitle

\noindent {\bf Introduction $\,$} Bose-Einstein condensates (BECs) have been created with the highly magnetic atoms  chromium \cite{Griesmaier2005a,*Bismut2010a}, dysprosium \cite{Mingwu2011a,*Lu2012a}  and erbium  \cite{Aikawa2012a}. The key new feature of these systems is a long ranged and anisotropic dipole-dipole interaction (DDI), which gives rise to a rich array of new physics for ultra-cold gases \cite{Lahaye_RepProgPhys_2009}. An important prediction made in 2003 was that a rotonic-like excitation would emerge in a condensate that is tightly confined along the direction that the dipoles are polarized \cite{Santos2003a} (also see \cite{Ronen2007a}).  To date there has been no experimental evidence for the existence of these excitations, and increasing attention is turning to developing signatures for detecting their presence, such as structure formation in the ground state density profiles \cite{Ronen2007a,Lu2010a,Wilson2012a,Martin2012a}, anisotropic superfluidity \cite{Wilson2010a,Ticknor2011a} and sensitivity to perturbations \cite{Wilson2008a,Nath2010a,Blakie2012a}.

The rotons within dipolar BECs arise when tight axial confinement, of length scale $l_z$, causes the effective interaction in the weakly confined (in-plane) directions to become momentum dependent \cite{Fischer2006a}. For in-plane momenta $\hbar k_{\rho}\lesssim\hbar/l_z$  the DDI is repulsive, while for $\hbar k_{\rho}\gtrsim\hbar/l_z$ it crosses over to being attractive, energetically softening  these modes and causing the roton to emerge with a wavelength $\lambda_{\mathrm{rot}}\!\sim\!l_z$. These rotons are sensitive to the condensate density and, due to the radial trapping, they are effectively confined as a \textit{roton gas}  in the high density central part of the condensate \cite{JonaLasinio2013}.

Here we investigate how rotons manifest in the fluctuations of a dipolar BEC. This is motivated by significant developments made in BEC imaging systems to allow precise \textit{in situ} measurements of density (e.g.~see \cite{Hung2011}). Such measurements actually determine the atom number $\hat{N}_\sigma$ within a finite-sized cell $\sigma$, with the cell parameters determined by the imaging system \cite{Hung2011a}. In equilibrium the number fluctuations 
 \begin{align}
  \delta N_\sigma^2 &\equiv \langle \hat{N}_\sigma^2\rangle -  {N}_\sigma ^2,\label{Eq:NumFluc}
 \end{align}
 about the mean number $N_\sigma\equiv\langle \hat{N}_\sigma\rangle$  are 
crucially dependent on the cell size \cite{Klawunn2011}: for cells smaller than all relevant length scales (i.e.~healing length and thermal de Broglie  wavelength) the fluctuations are Poissonian  $  \delta N_\sigma^2=  {N}_\sigma$, whereas for large cells   the thermodynamic result holds $\delta N_\sigma^2={N}_\sigma k_BTn\kappa_T$, with $n$ the density and $\kappa_T$ the isothermal compressibility. DDIs introduce two important features for fluctuations: (1) the roton wavelength introduces a new length scale which is revealed as the cell size changes; (2) the attractive nature of the interaction for short wavelengths induces enhanced fluctuations, this is in contrast to the suppression of fluctuations expected  in a condensate with short-ranged $s$-wave interactions \footnote{Large  BECs with purely contact interactions are only stable when $a_s>0$, which suppresses long wavelength fluctuations.}.

 \begin{figure}[H]
\begin{center}
\includegraphics[width=3.1in]{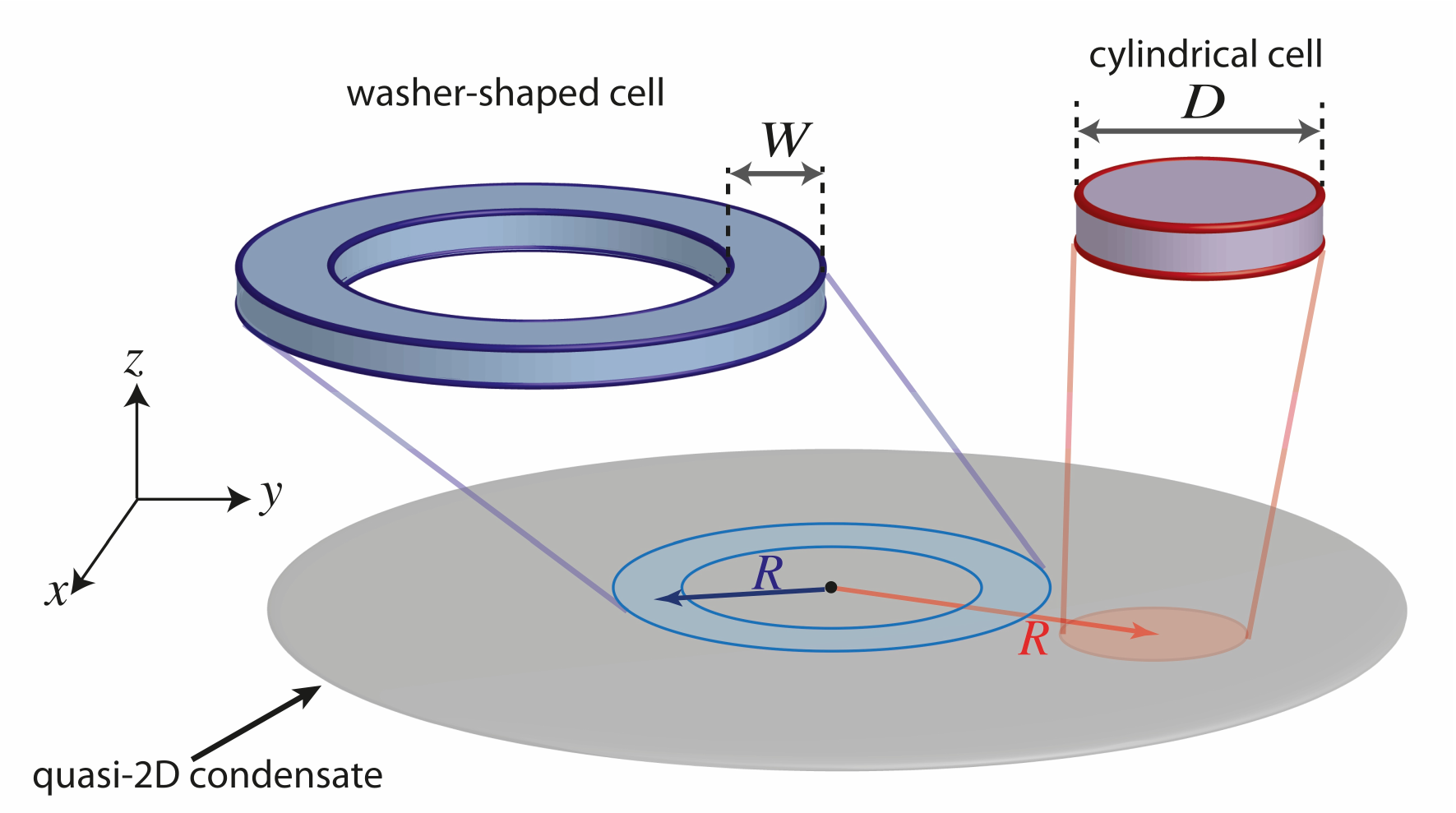}
\caption{(color online) Schematic   geometry of the system we consider: a radially symmetric pancake condensate realized by tight confinement along $z$; cylinder and washer-shaped cells in which  number fluctuations are measured. The cylindrical cell is parameterized by the radius $R$ from trap center to the cylinder axis, and its diameter $D$. The washer-shaped cell is centered on the trap center, and is parameterized by the mid-radius   $R$ and  width $W$. For both cells the $z$ extent is taken to be larger than the condensate thickness. 
 \label{Fig:Geometry}}
\end{center}
\end{figure}

\noindent {\bf Formalism $\,$} 
The geometry of the tightly confined (pancake) dipolar condensate and the types of cells in which we measure number fluctuations  are shown in  Fig.~\ref{Fig:Geometry}. The two cell shapes we consider are cylindrical and washer-shaped cells with the $z$ extent of these cells being larger than the condensate thickness in this direction.  The cylindrical cell can be realized experimentally using absorption imaging (column density along $z$) with the minimum diameter ($D$)  being set by the spot size of the imaging system (e.g.~see \cite{Trebbia2006a,Jacqmin2011a,Hung2011a}). These measurements are made using a charged coupled device in which the pixel size is typically much smaller than the spot size. Cells with more general shapes in the $xy$-plane can be realized by collecting the signal from a set of pixels, e.g.~larger diameter cylinders and the washer-shaped case that we consider later.

The cell number operator is   $\hat{N}_\sigma \equiv \int_\sigma d^3\x\,\hat\Psi^\dagger(\x)\hat\Psi(\x)$, with $\hat\Psi(\x)$ the bosonic field operator and $\sigma$ denoting the cell volume. 
Within a meanfield approach, valid for weakly interacting condensates, the field operator takes the form  
 \begin{equation}
 \hat\Psi(\x) \approx \psi_0(\x) + \sum_i\left[u_i(\x)\hat\alpha_i - v_i^*(\x)\hat\alpha_i^\dagger\right].\label{psihatbog}
\end{equation}
In the above expression   $\psi_0$ is the condensate wavefunction (normalized to the condensate number $N_0$), which satisfies the nonlocal dipolar Gross-Pitaevskii equation (GPE) \cite{Goral2000a}
\begin{equation} 
\mu\psi_0 =\left[\!-\frac{\hbar^2\nabla^2}{2m}\!+\!V(\x)\!+\!\int d^3\x^\prime U(\x\!-\!\x^\prime)| \psi_0(\x^\prime)|^2\right]\psi_0 ,
\end{equation}
where  $\mu$ is the chemical potential.  For dipoles  polarized along $z$  the inter-atomic interaction potential is of the form
\begin{equation} 
U(\mathbf{r}) =g\delta(\mathbf{r})+ \frac{3g_{\rm dd}}{4\pi}\frac{1-3\cos^2\theta}{|\mathbf{r}|^3},
\end{equation} 
where the short-range interaction is characterized by the contact parameter $g=4\pi a_s\hbar^2/m$, with $a_s$ the $s$-wave scattering length. The DDI parameter is   $g_{\rm dd} = \mu_0\mu_m^2/3$, with $\mu_m$ is the magnetic dipole moment, and $\theta$ is the angle between $\mathbf{r}$ and the $z$ axis.  
The atoms are taken to be harmonically confined by a potential $V(\x) = \frac{1}{2}m\omega_\rho^2(\rho^2+\lambda^2z^2)$ with  $\lambda=\omega_z/\omega_\rho$.
The amplitudes $\{u_i,v_i\}$, with energy eigenvalues $\epsilon_j$, describe the  linearized quasiparticle excitations, and  are obtained by solving nonlocal Bogoliubov de Gennes equations \cite{Ronen2006a}. The quasiparticle operators satisfy $[\hat{\alpha}_i,\hat{\alpha}_j^\dagger]=\delta_{ij}$. Our results should be quantitatively accurate as the maximum depletion is small ($<7\%$ of $N_0$). We also note that meanfield theory has been shown to accurately describe the stability boundaries found in experiments and associated dynamics \cite{Wilson2009B,Koch2008,Muller2011}.

The mean cell number is $N_\sigma=\int_\sigma d^3\x\,n(\x)$, 
 where $n(\x)=|\psi_0(\x)|^2+\tilde{n}(\x)$ is the total density, with
\begin{equation}\tilde{n}(\x)=\sum_j\left[\bar{n}_j|u_j(\x)|^2+(\bar{n}_j+1)|v_j(\x)|^2\right],\end{equation} 
 the depletion (noncondensate) density,  $\bar{n}_j =\left(e^{\beta\epsilon_j}-1\right)^{-1}$ and $\beta=1/k_BT$   the inverse temperature. Neglecting terms greater than second order in the quasiparticles, we obtain
 \begin{align}
 \delta N_\sigma^2 &=  \!\sum_i\!\int_\sigma d^3\x_1\!\int_\sigma d^3\x_2 \,\delta n_i(\x_1)\delta n_i^*(\x_2)\coth\left(\tfrac{1}{2}\beta\epsilon_i\right),\label{dN2}
\end{align}
with $\delta n_i \equiv \psi_0\left(u_i -v_i \right)$ the density fluctuation amplitude of the $i$th quasiparticle, where we have taken $\psi_0$ to be real.

 \begin{figure}[!tbh]
\begin{center}
\includegraphics[width=3.15in]{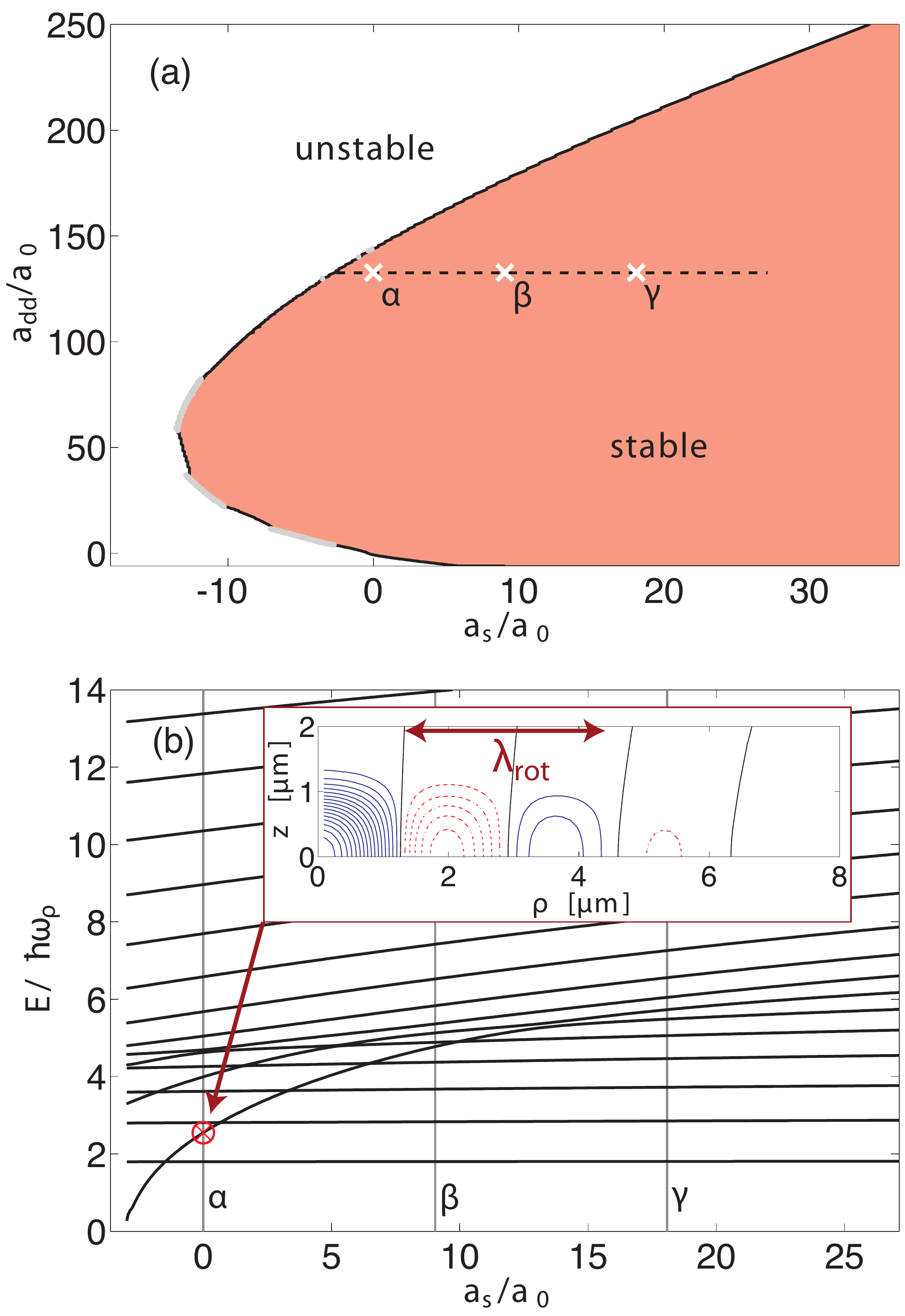}
\caption{(color online) (a) Stability phase diagram in interaction parameter space for  $N_0=2.5\times10^4$ $^{164}$Dy atoms in a $\lambda=20$ trap with $\omega_\rho= 2\pi\times 10.8\,$s$^{-1}$. 
 The shaded region indicates where a stable solution exists and the gray portions of the boundary mark the boundary of density oscillation islands \cite{Ronen2007a}. 
Crosses indicate the three sets of parameters we examine in this paper  with $a_s/a_0=0,9.04, 18.1$ as the cases we label as $\alpha,\beta,\gamma$, with $a_0$ the Bohr radius. 
The dipole length is $a_{\mathrm{dd}}  \equiv m {g_{\mathrm{dd}}}/{4\pi\hbar^2}\approx132.6\,a_0$ for $^{164}$Dy.
(b) Quasiparticle energies of  $m_z=0$ modes  as $a_s$ is varied  along the dashed path in (a). Vertical lines mark  the cases $\alpha,\beta$ and $\gamma$, and the crossed circle indicates the lowest energy roton mode.The contours of the density fluctuation  $\delta n_i$ for the roton shown in inset. Contour spacing is $3\times10^{11} $cm$^{-3}$, solid lines (dot-dashed) represent positive (negative) contours, while nodes are shown as thick black lines.
 \label{Fig:PhaseDia}}
\end{center}
\end{figure}
In Fig.~\ref{Fig:PhaseDia}(a) we show the numerically determined stability diagram for $N_0=2.5\times10^4$ $^{164}$Dy atoms in a trap of aspect ratio   $\lambda=20$, indicating the  range of contact and dipolar interactions where all the quasiparticle energies are real and positive. In the upper part of the stable region, where the DDI dominates, roton excitations tend to develop (e.g.~see \cite{Blakie2012a}). Indeed, in Fig.~\ref{Fig:PhaseDia}(b) we show the energy spectrum along the dashed path in Fig.~\ref{Fig:PhaseDia}(a) where we vary $a_s$, as can be done using a Feshbach resonance. As $a_s$ becomes smaller we observe a radially excited roton mode soften,  leading to  instability  for $a_s\approx-3a_0$.  We note that there is a range of roton modes softening, including those with nonzero values of angular momentum projection $m_z$ (which are not shown in the spectrum)  although the modes that soften all have a similar wavelength, as indicated in the inset to  Fig.~\ref{Fig:PhaseDia}(b). In Fig.~\ref{Fig:PhaseDia} we  indicate three parameter sets that we will use here  and  have labeled $\alpha,\beta$ and $\gamma$ (differing by the value of $a_s$). The purely dipolar case $\alpha$ has the most prominent rotons, while cases $\beta$ and $\gamma$ (progressively further from the stability boundary) may be more convenient to explore experimentally.

\begin{figure} [!tbh]
\begin{center}
\includegraphics[width=3.3in]{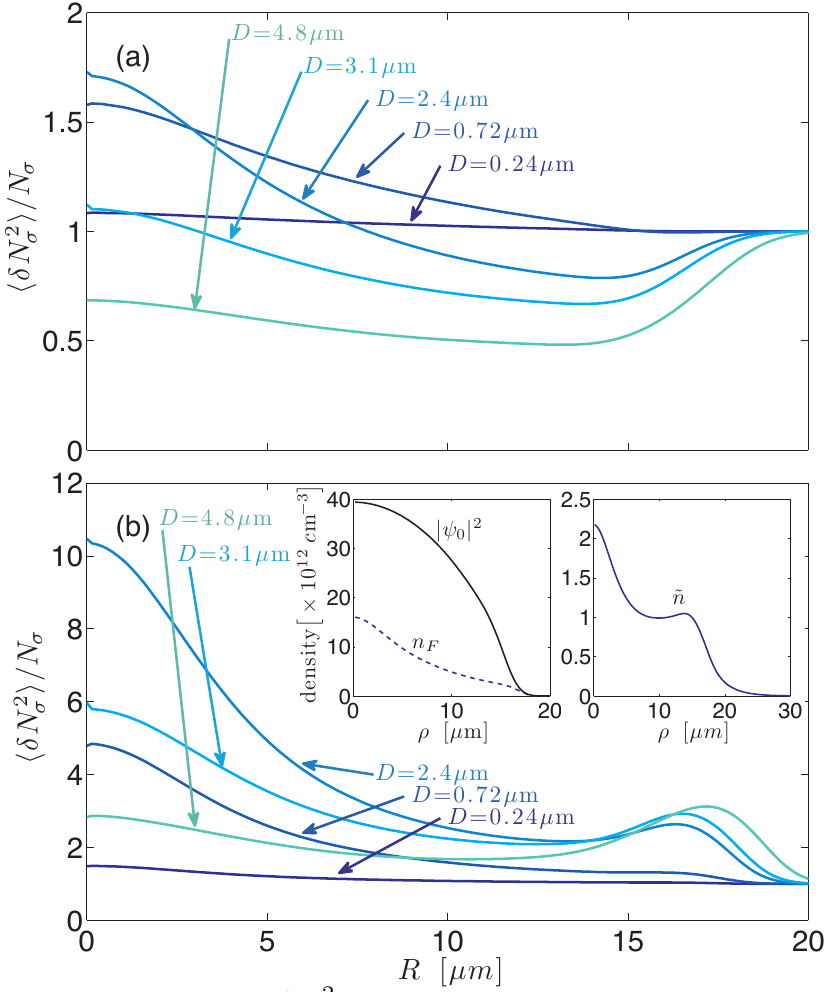}
\caption{(color online) $\delta N_\sigma^2$ as a function of cylinder position $R$ for various diameters $D$ for case $\alpha$. 
 (a) $T = 0$ with a depletion of $\tilde{N}=171$. (b) $T = 5.2\,$nK with $\tilde{N}=1570$. Insets: Condensate,  depletion and fluctuation density. Densities evaluated at $z=0$ and  $n_F$ is calculated using modes with $\epsilon_j<20\hbar\omega_\rho$.\label{Fig:Disks}}
\end{center}
\end{figure}

\noindent {\bf Cylindrical Cells $\,$}  We present results for $\delta N_\sigma^2$ for the purely dipolar case $\alpha$    in Fig.~\ref{Fig:Disks}. These results are calculated using Eq.~(\ref{dN2}) in normally ordered form and include contributions from $\sim10^4$ quasiparticle modes.  We emphasize that our calculations are fully three-dimensional and we have performed tests with up to $10^5$ modes finding changes in $\delta N^2_\sigma$ of $\lesssim$0.7\%. The total depletion is $\tilde{N}=\int d^3\x\,\tilde{n}\approx 1570$ (i.e.~$6\%$ of $N_0$) at the highest $T$ considered.
The peak in the depletion density [see inset to Fig.~\ref{Fig:Disks}(b)] near trap center reveals the presence of the  confined roton modes. To emphasize that the rotons have a larger effect on the fluctuations than is suggested by the depletion, we define and calculate a \textit{fluctuation density} by the correlator $n_{ {F}}(\x) \equiv \sqrt{G^{(2)}(\x)-n(\x)^2}$, where $G^{(2)}\equiv\langle(\hat{\Psi}^\dagger)^2(\hat{\Psi})^2\rangle$  (cf.~quasicondensate definition in \cite{Prokofev2001a}). While $n_F$ is generally less than $\tilde{n}$ in a repulsive condensate with contact interactions, for the dipolar gas it is significantly enhanced [see inset to Fig.~\ref{Fig:Disks}(b)] \cite{Boudjemaa2013a}. For this reason a self-consistent treatment of the excitations (e.g.~see \cite{Ronen2007b,Ticknor2012a}) in this regime would be desirable.

At $T=0$ [Fig.~\ref{Fig:Disks}(a)] and for $R=0$ (cells at trap center) the number fluctuations are almost Poissonian  ($\delta N_\sigma^2/N_\sigma\approx 1$)  for the smallest cell diameter $D = 0.239~\mu$m, increase to become super-Poissonian for $D \sim 2~\mu$m, and then decrease to sub-Poissonian for  $D \gtrsim3~\mu $m.
The dominant contribution to the fluctuations in Eq.~(\ref{dN2}) arises from modes with wavelength comparable to the cell width \cite{Klawunn2011}. The dominant modes for the smallest size cells are the high energy (free-particle-like) excitations, while for the largest cells it is the long wavelength repulsively interacting phonon-like modes that suppress fluctuations. For intermediate cell sizes comparable to half the roton wavelength $\lambda_{\mathrm{rot}}/2\approx1.5\,\mu$m [see inset to Fig.~\ref{Fig:PhaseDia}(b)] the fluctuations are enhanced by the effective attractive  character of the DDI  for modes at this scale. This  \textit{roton peak} in $\delta N_\sigma^2$ is a key feature of dipolar condensates and can be used as a signature of the emergence of roton excitations.

Temperature plays an important role in BEC fluctuations, and  the  results in Fig.~\ref{Fig:Disks}(b) are for a case typical of the lowest temperatures obtained in experiments   ($\approx\!13\%$ of the critical temperature) \footnote{We restrict our attention to $T\ll T_c$ and neglect the interaction of excitations with the thermal component, c.f.~\cite{Ronen2007b,Bisset2012,Ticknor2012a}}. 
These results show that thermal effects wash out the sub-Poissonian fluctuations in large cells, while the  {roton peak} in $\delta N_\sigma^2$ (occurring for $D\sim\lambda_{\mathrm{rot}}/2$) is significantly enhanced.
 For both temperature regimes in Fig.~\ref{Fig:Disks} the height of the roton peak decreases with $R$ as the cell is moved away from the central region, because  the rotons are effectively confined to the trap center \cite{JonaLasinio2013}.

\begin{figure}[!tbh]
\begin{center}
\includegraphics[width=3.4in]{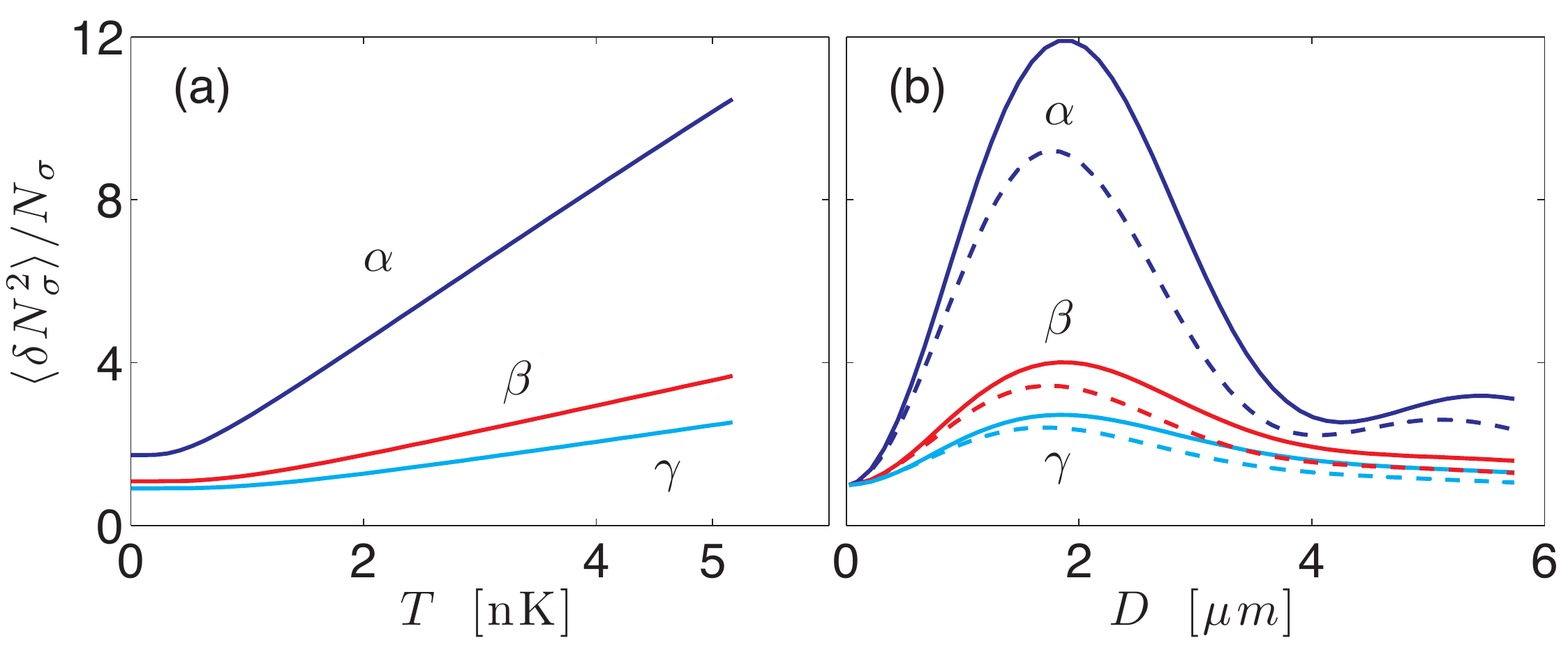}
\caption{(color online) Peak fluctuations (at $R=0$) within a cylindrical cell  (a) as  a function of temperature at diameter $D=1.67\,\mu$m, and (b) as a function of  cylinder diameter at $T = 5.2\,n$K. Each graph has results for three different values of   contact interaction  (see Fig.~\ref{Fig:PhaseDia}). Solid: full calculations and dashed: LDA result (see text). 
 \label{Fig:TempAna}}
\end{center}
\end{figure}

Figure~\ref{Fig:TempAna}(a) examines the fluctuations at  $R=0$  for a cylinder of fixed diameter  $D=1.67~\mu$m, chosen to be sensitive to the roton excitations, as a function of temperature and for several values of the contact interaction. 
In all cases the plateau at low $T$ represents the quantum fluctuations associated with the rotons, while the transition to a rising line coincides with the thermal activation of these modes, occurring when $k_BT$ is of the order of the lowest roton energy. With reference to spectra in Fig.~\ref{Fig:PhaseDia}(b) we note that the energy of the lowest roton increases as $a_s$ increases.    In Fig.~\ref{Fig:TempAna}(b) we show the dependence of $\delta N_\sigma^2$ on cell diameter.
Importantly, for the results with larger $a_s$  (where the roton energy is higher), the roton peak in $\delta N_\sigma^2$ at $D\sim\tfrac{1}{2}\lambda_{\mathrm{rot}}$ is smaller, but still discernible.
We emphasize that the fluctuation signal for case $\beta$  peaks at a value of 4 times the Poissonian case.

In Fig.~\ref{Fig:TempAna}(b)  we also present results of  a simplified theory based on the local density approximation (LDA) for calculating the fluctuations in the quasi-two-dimensional (quasi-2D) regime, valid when $k_BT<\hbar\omega_z$. This LDA theory makes use of the GPE wavefunction but avoids the computationally challenging step of calculating all the quasiparticles and integrating them over the cell region $\sigma$, as required to use Eq.~(\ref{dN2}).
We first note that for a uniform quasi-2D system the fluctuations in a disk of radius $D$ is 
 \begin{equation}
 \delta N_\sigma^2 = n_{2D} \int  d^2\mathbf{k}_{\rho}\frac{J_1^2(\tfrac{1}{2}Dk_{\rho})}{k^2_{\rho}}S(\mathbf{k}_{\rho}) ,\label{Eq:NumFlucUni}
 \end{equation}
 with $S(\mathbf{k}_{\rho})$ the static structure factor, $n_{2D}$ the areal density, and the Bessel function term arising as a geometric factor for the cell  \cite{Klawunn2011}.
\begin{figure}[h]
\begin{center}
\includegraphics[width=3.2in]{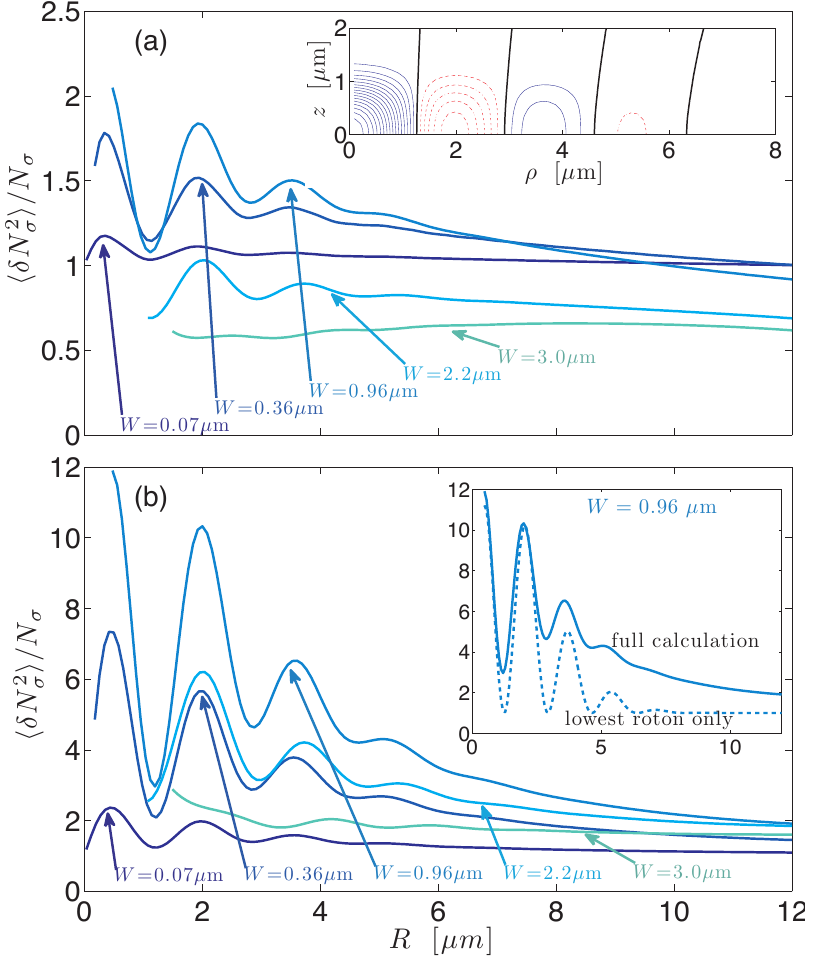}
\caption{(color online) Fluctuations versus washer radius $R$ and for various washer widths $W$ for the $a_s=0$. (a) $T=0$ results. Inset:   $\delta n_i$  for the lowest   roton mode [repeated from inset of Fig.~\ref{Fig:PhaseDia}(b)].\\ (b) $T = 5.2\,n$K results. Inset: The dashed curve shows the number fluctuations calculated using only  the lowest roton mode, compared to the full calculation (solid).
 \label{Fig:Annuli}}
\end{center}
\end{figure}

The basis of our approach for extending this result to the trapped system is  to use $\psi_0$  to calculate a \textit{local} Bogoliubov dispersion relation $\epsilon_{\mathrm{B}}( \mathbf{k}_\rho,\bm\rho)$ at each radial position $\bm\rho=(x,y)$ (see  \cite{Blakie2012a} for details), from which a local static structure factor can be obtained as $S(\mathbf{k}_{\rho},\bm\rho)=\frac{\hbar^2k_{\rho}^2}{2m\,\epsilon_B}\coth
\left( \tfrac{1}{2}\beta\epsilon_{\mathrm{B}}\right).$ The static structure factor for the system within the cell $\sigma$ can then be computed by the density weighted LDA sum    $S_{\sigma}(\mathbf{k}_{\rho})\!=\!\int_{\sigma} d^2\bm{\rho}\,n_{2D}(\bm{\rho})  S(\mathbf{k}_{\rho},\bm{\rho})$, where  $n_{2D}(\bm{\rho})\equiv\int dz\,|\psi_0(\bm{\rho},z)|^2$. The number fluctuations for the trapped system is then given by Eq.~(\ref{Eq:NumFlucUni}) with the replacement $S(\mathbf{k}_{\rho})\to S_{\sigma}(\mathbf{k}_{\rho})$. The comparison in Fig.~\ref{Fig:TempAna}(b) demonstrates that the LDA results provide a useful quantitative estimate.

\noindent {\bf Washer-Shaped Cells $\,$}  It is interesting to consider more general cells shapes for measuring $\delta N_\sigma^2$. Here we examine the nonsimply connected  washer-shaped cell (see Fig.~\ref{Fig:Geometry}).  We show $\delta N_\sigma^2$ for this cell versus washer radius $R$ and for various washer widths at both  zero [Fig.~\ref{Fig:Annuli}(a)] and finite temperature [Fig.~\ref{Fig:Annuli}(b)]. The most striking feature of these results is the emergence of fine oscillations in $\delta N_\sigma^2$ as $R$ is varied. This arises because the washer geometry only allows $m_z=0$ modes to contribute \footnote{Selectivity to $m_z\!=\!0$ modes arises in  (\ref{dN2}) because the washer cell is radially symmetric and the azimuthal character of the $\delta n_i$ fluctuations for $m_z\ne0$ integrate to zero.}, thus only a small number of modes contribute to the result. Indeed, in the inset to  Fig.~\ref{Fig:Annuli}(b) we compare the full result for $\delta N_\sigma^2$ to that calculated by restricting the sum in Eq.~(\ref{dN2}) to just the lowest energy roton mode, and find reasonable agreement, with the difference being due to other excited $m_z=0$ modes.  The broad envelope of these oscillations is similar (as a function of $R$) to that observed for the cylindrical cells in Fig.~\ref{Fig:Disks} (reflecting roton confinement). Further refinement of the cell geometry (e.g.~segments of a complete washer) could be used to select for $m_z\ne0$ roton modes which are predicted to play an important role in biconcave ground states (see \cite{Ronen2007a}).
 
 \noindent {\bf Conclusions $\,$}
In this paper we have examined the number fluctuations that occur within  small cells of a dipolar BEC. We have shown that these fluctuations provide a large signal indicating the presence of roton excitations, even when the depleted density associated with these excitations is small. 
 Our results show that the fluctuations are highly dependent on the cell size relative to the roton wavelength, and should be practical to measure in experiments with  imaging  that can resolve cells of size   $\sim\!2~\mu$m.  This resolution has been approached in similar style quasi-2D experiments with Cs  \cite{Hung2011a}, and  well exceeded with other quantum gas microscope (e.g.~\cite{Bakr2009}) and in scanning electron microscopy  \cite{Gericke2008a} experiments.
 While we have focused on a physically realizable regime with Dy, we emphasize that Cr and Er systems are also suited to this type  of measurement.

\noindent {\bf Acknowledgments $\,$}  This work was supported by the Marsden Fund of New Zealand (contract UOO0924).



%

\end{document}